\documentclass{elsart}
\usepackage{graphicx}
\usepackage{epsfig}
\usepackage{units}
\usepackage{amssymb}
\begin{document}

\begin{frontmatter}



\title{Growth and magnetism of self-organized arrays of Fe(110) wires formed by deposition on kinetically grooved W(110)}


\author{B. Borca, O. Fruchart, F. Cheynis, M. Hasegawa, C. Meyer}

\address{Laboratoire Louis N\'eel, Grenoble, France}

\begin{abstract}
Homoepitaxy of W(110) and Mo(110) is performed in a kinetically-limited regime to yield a
nanotemplate in the form of a uniaxial array of hills and grooves aligned along the [001]
direction. The topography and organization of the grooves were studied with RHEED and STM. The
nanofacets, of type \{210\}, are tilted $\sim18^\circ$ away from (110). The lateral period could
be varied from 4 to \unit[12]{nm} by tuning the deposition temperature. Magnetic nanowires were
formed in the grooves by deposition of Fe at $150^\circ$C on such templates. Fe/W wires display an
easy axis along [001] and a mean blocking temperature $T_{B}\approx\unit[100]{K}$.
\end{abstract}


\end{frontmatter}

% main text
\section{Introduction}
\label{intro}

Over the past decades magnetic thin films have been the focus of intense work. When their
thickness is reduced surface and dimensionality effects become dominant, affecting e.g. magnetic
anisotropy and thermal excitations. More recently the focus has been extended to systems of lower
dimensionality , e.g. nanosized wires and dots. Beyond the limits of chemical routes \cite{Fer99}
and lithography \cite{Mar03}, self-organization is the most suited technique to achieve
nanostructures down to the atomic size \cite{Fru05}. Self-organized dots \cite{Roh06,Voi91} and
wires \cite{Gam03} with a long range order could be grown on templates like surface
reconstructions and vicinal surfaces. Achieving self-organization on surfaces that spontaneously
do not form a template is appealing, because it should be more versatile. Concerning wires
fabricated on \textsl{non-vicinal} surfaces, one route is to ion-etch continuous film under
grazing incidence \cite{Val02}. Here we propose an alternate route, consisting in growing magnetic
wires on a non-magnetic nanotemplate of grooves, which is achived by a growth process
out-of-equilibrium.
% \appendix

\section{Experimental procedures}
% \label{}
The samples were grown in an ultra-high vacuum chamber using pulsed-laser deposition with a Nd-YAG
laser at \unit[532]{nm}. The chamber is equipped with a quartz microbalance, sample heating and a
translating mask for the fabrication of wedge-shaped samples. We use a 10 keV Reflection High
Energy Electron Diffraction (RHEED) setup with a CCD camera synchronized with laser shots, which
permits operation during laser operation and thus deposition. An Omicron room-temperature Scanning
Tunneling Microscope (STM), used in the constant current mode, and  an Auger Electron Spectrometer
(AES) are located in connected chambers. The metallic films are grown on commercial sapphire
single crystals. A detailed description of the chamber and growth procedures can be found in
\cite{Fru}. The magnetic measurements were performed ex-situ on capped samples, with a
Superconducting QUantum Interference Device (SQUID) magnetometer.

\section{Growth and structural investigations}
\subsection{Growth of a self-organized template}

\begin{figure*}
  \center
  \includegraphics[width=140mm]{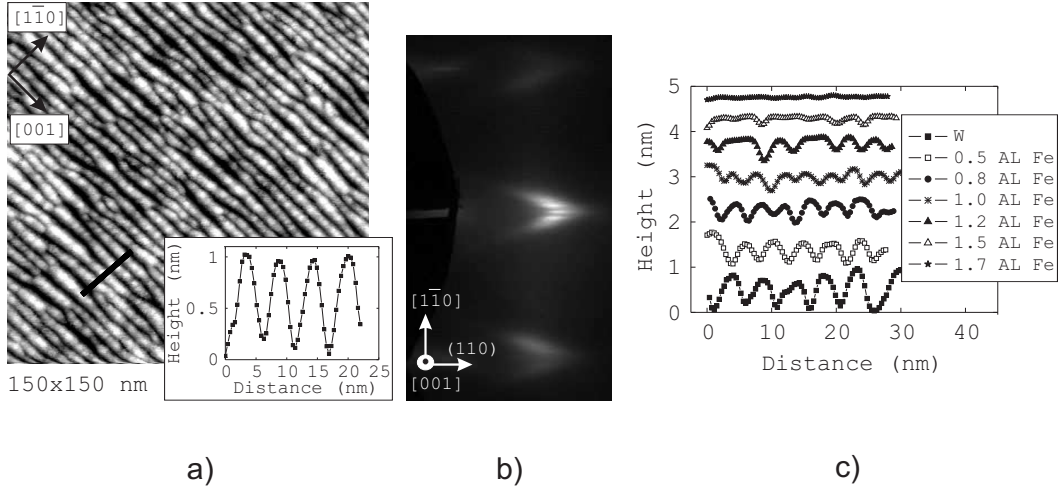}
  \caption{a) STM image of the W(110) surface with uniaxial hills and grooves along the [001] direction, prepared by deposition
   of \unit[10]{nm} at $150^\circ$C on a flat W surface. Inset: cross-section along the black line; b) RHEED pattern along the [001]
    azimuth on the surface shown in a). The sample is vertical on the left side of the pattern; c) Cross sections across the wires
     for different nominal thicknesses of Fe, listed in atomic layers (AL)}
  \label{fig:Graphique}
  %\label{fig:RheedSGr}
  %\label{fig:CrossFe}
\end{figure*}

We start from Mo(110) or W(110) buffer layers (thickness $\sim\unit[10]{nm}$) deposited on
Al$_{2}$O$_{3}$ (11$\overline{2}$0) substrates. The process consists of deposition at room
temperature (RT) followed by annealing at $800^\circ$C, yielding a surface of quality similar to
that of metal single crystals \cite{Fru98}. In the case of W an underlayer of Mo(\unit[1]{nm}) is
deposited first at RT, to avoid twinning of the annealed film \cite{Fru}.

It has been reported upon deposition at moderate temperature along the (110) plane, that Fe
\cite{Alb93} and W \cite{Koh00} develop a self-organized surface with uniaxial hills and grooves
oriented along the [001] direction. The origin of this structure was explained by an anisotropic
diffusion at the nucleation stage, followed by a kinetic effect related to the Ehrlich-Schwoebel
barrier. The average angle of the grooves was reported and simulated to increase with nominal
thickness and with lowering temperature. No limiting angle was discussed in \cite{Alb93}, while in
\cite{Koh00} a limiting angle of $26.57^\circ$ associated with \{310\} facets was predicted.

Here we confirm the formation of grooves for W (fig.1-2) and report it also for Mo. We focus on
deposition at moderate temperature for which the angle of the facets is significant, i.e.
$\leq250^\circ$C for Mo and $\leq350^\circ$C for W. Monitoring RHEED during growth we observe that
the angle of the facets becomes stationary after a while, typically a few nanometers of nominal
thickness. RHEED and STM show in agreement that the stationary angle is close to $\pm18^\circ$,
consistent with facets of type \{210\} (fig.1-2). Thus the prediction of Ref. \cite{Koh00} for a
stationary angle is confirmed, however the resulting facet is not the one that was foreseen.

The quality of the order is reflected on the RHEED patterns by the occurrence of satellite streaks
(fig.1b), whose splitting agrees quantitatively with the period found by STM, e.g. \unit[5]{nm}
for deposition at $150^\circ$C (fig.1a). The average depth of the grooves is here \unit[1]{nm}. In
\cite{Alb93} it was argued that the period is selected during the nucleation stage in the
sub-atomic-layer range of deposition. We thus deposited \unit[15]{nm} of W under decreasing
temperature (from $550^\circ$C to $150^\circ$C) allowing us to reach a period of 10-\unit[12]{nm}
and grooves of depth 2-\unit[2.5]{nm}. STM and RHEED (fig.2a-b) reveal again the \{210\} faceting,
which becomes stationary after 1-\unit[2]{nm} have been deposited. Notice that the order is
slightly decreased for this process, with respect to deposition at $150^\circ$C.

\begin{figure*}
  \center
  \includegraphics[width=140mm]{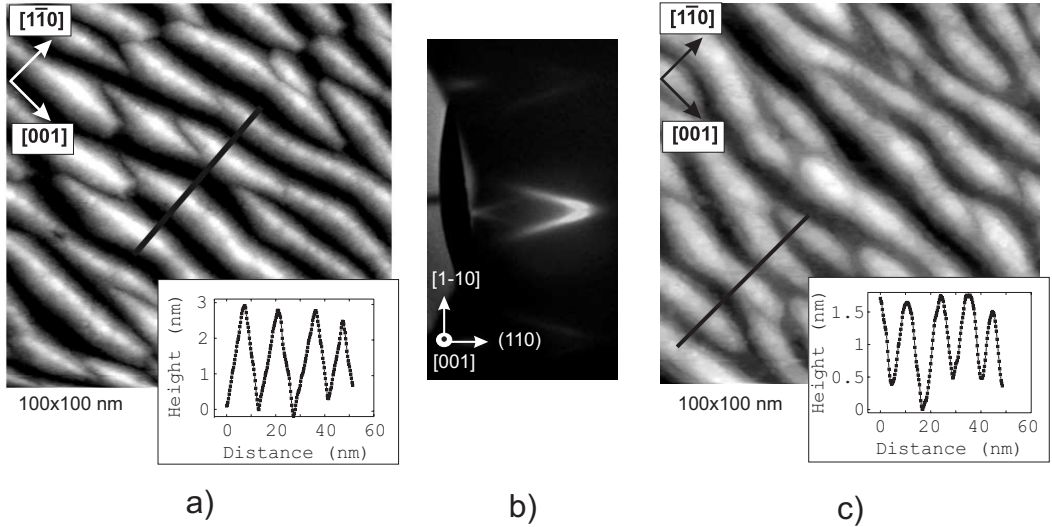}
  \caption{a) STM image and the cross-section of a W(110) template prepared by a deposition at a decreasing gradient of
   temperature ($550^\circ$C-$150^\circ$C) on a flat W(110) surface; b) RHEED pattern along the [001] azimuth for a W(110)
   surface as in a); c) \unit[2.5]{AL} of Fe , deposited at $150^\circ$C on a W template as in a)}
  \label{fig:Graph2}
  %\label{fig:RheedBGr}
  %\label{fig:BGrFe}
\end{figure*}

\subsection{Fabrication of nanowires}

We have used the surfaces with hills and grooves reported above as a template for the self-organization  of magnetic wires. We focus on W as a material template because the faceting remains stable until a higher temperature than in the case of Mo. We deposited Fe at $150^\circ$C for which Fe grows essentially layer-by-layer on \textsl{flat} W(110) \cite{Alb93}. Samples wedged from 0 (W template) to 3 atomic layers were investigated by STM. Cross-sections across the wires (fig.1c) and calculation of the roughness on the STM images show that the depth of the grooves decreases with the amount of Fe deposited. This suggests that Fe grows inside the grooves and not uniformly or at the top of the hills, and thus forms wires. This conclusion is currently under investigation by quantitative AES peak analysis and TEM. The process described here works equally well independently from the period of the template.

\section{Magnetic measurements}

\begin{figure*}
  \center
  \includegraphics[width=140mm]{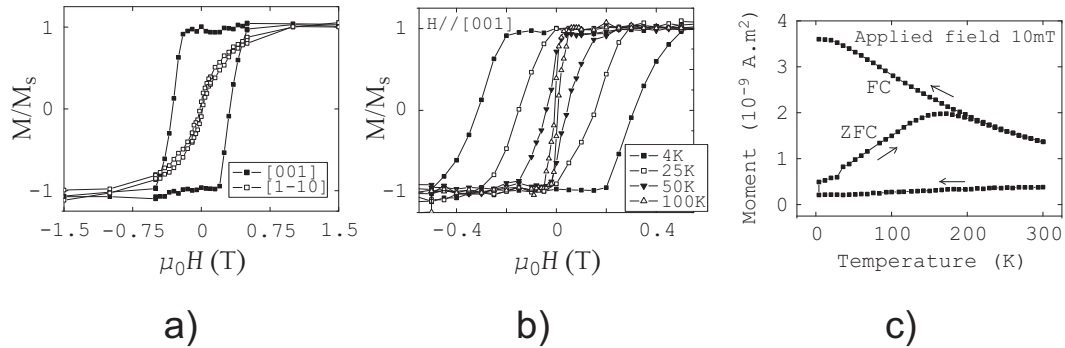}
  \caption{Magnetic results of a \unit[2.5]{AL} of Fe in the form of wires with a mean \unit[10]{nm} period, \unit[7]{nm} width,
  and \unit[1]{nm} height. a) Hysteresis loops at \unit[10]{K}, with an in-plane applied field parallel and perpendicular
  to [001] direction; b) Hysteresis loops as a function of temperature in an applied magnetic field parallel to [001] direction;
  c) Variation of the low-field magnetization as a function of temperature within a "zero-field cooling/field cooling" process.
  The applied field is \unit[10]{mT} }
  \label{fig:Graph3}
  %\label{fig:Measy}
  %\label{fig:MT}
\end{figure*}
We focus here on Fe wires prepared on W(110) templates with a mean period of \unit[10]{nm}. For an Fe nominal thickness of \unit[2.5]{AL} the wires are expected to have a mean width of \unit[7]{nm} and a mean height of \unit[1]{nm} from the topography observed by STM. The magnetic properties are examined \textsl{ex-situ} by SQUID using a \unit[5]{nm}-Mo-capped sample. Hysteresis loops were performed along two in-plane directions: parallel to Fe[001], i.e. along the wires and parallel to Fe[1$\overline{1}$0], i.e. across the wires. The easy magnetization axis is [001](fig.3a). The anisotropy energy is calculated along the [1$\overline{1}$0] loop like $E_{\mathrm{a}}=\mu_{0}\int^{M_{S}}_{0}H dM$. This anisotropy originates as the sum of many contributions, among which only the dipolar energy can be estimated reliably: $E_{\mathrm{d}}\approx3\times\unit[10^{5}]{J/m^{3}}$. Other sources of anisotropy are surface (N\'eel-type) expected to favor the [1$\overline{1}$0] direction for the (bottom) Fe/W interface \cite{Gra86} and [001] for the (top) Mo/Fe interface \cite{Fru99}, step-edge for Fe/W, expected to favor [1$\overline{1}$0] \cite{Hau98}; magneto-elastic -- unlike thin films, investigated so far, stress is both in-plane and out-of plane owing to the occurrence of many steps, implying significant shear, so that no figure or even sign can be reliably predicted; finally the bulk Fe anisotropy is negligible. Surprisingly $E_{\mathrm{a}}\approx E_{\mathrm{d}}$ despite this complex situation, similarly to (Fe,Ag) self-organized arrays of wires \cite{Bor06}. \\
The loops measured with the external field H//[001] at different temperatures have a rather square shape, and at remanence full saturation is still observed (fig.3b). The coercivity continuously decreases with temperature while M$_{\mathrm{S}}$ remains essentially unchanged suggesting a superparamagnetic behavior.The ultimate blocking temperature determined by the "zero field cooling/field cooling" process is $T_{B}\approx \unit[160]{K}$ (fig.3c). The \textsl{mean} value of T$_{B}$ is around \unit[100]{K}, estimated as the half-way of the zero-field cooling remagnetization curve.

\section{Conclusion}

Arrays of Fe nanowires were obtained by deposition on self-organized kinetically faceted surfaces of W, with a period in the range 4-\unit[12]{nm}. Arrays with a mean period of \unit[10]{nm}, prepared by deposition of an amount of \unit[2.5]{AL} of Fe, display an in-plane uniaxial anisotropy along [001], the direction of the wires. The wires are superparamagnetic at room temperature and the mean blocking temperature is \unit[100]{K}. Efforts to rise this blocking temperature are under way.

\section*{Acknowledgements}

We are grateful to Ph. David and V. Santonacci for technical support.

\end{document}